\documentclass[aps,prr,twocolumn,showpacs,superscriptaddress]{revtex4-1}
\usepackage{amsmath,amssymb}
\usepackage{graphics,graphicx,color}
\usepackage{dcolumn,bm}
\usepackage[normalem]{ulem}
\usepackage{tikz}
\usepackage{soul}
\usepackage[colorlinks=true,citecolor=blue,urlcolor=blue]{hyperref}

\def\[{\left[}
\def\]{\right]}
\def\({\left(}
\def\){\right)}
\def\be{\begin{equation}}
\def\ee{\end{equation}}
\def\bea{\begin{eqnarray}}
\def\eea{\end{eqnarray}}

\newcommand{\gaug}
{\affiliation{Institute for Theoretical Physics, Georg-August-Universit\"at G\"ottingen, 37077 G\"ottingen, Germany}}

\begin{document}
\title{Memory in Non-Monotonic Stress Response of an Athermal Disordered Solid}

\author{Rituparno Mandal}%
\email[Email: ]{rituparno.mandal@uni-goettingen.de}
\gaug

\author{Diego Tapias}
\email[Email: ]{diego.tapias@theorie.physik.uni-goettingen.de}
\gaug

\author{Peter Sollich}%
\email[Email: ]{peter.sollich@uni-goettingen.de}
\gaug
\affiliation{Department of Mathematics, King's College London, London WC2R 2LS, UK}

\begin{abstract}

Athermal systems across a large range of length scales, ranging from foams and granular bead packings to crumpled metallic sheets, exhibit slow stress relaxation when compressed. Experimentally they show a non-monotonic stress response when decompressed somewhat after an initial compression,  {\textit{i.e.}}\ under a two-step, Kovacs-like protocol. It turns out that from this response one can tell for how long the system was in a compressed state, suggesting an interpretation as a memory effect. In this work we use a model of an athermal jammed solid, specifically a binary mixture of soft harmonic particles, to explore this phenomenon through {\textit{in-silico}} experiments.  Using extensive simulations under conditions analogous to those in experiment, we observe identical phenomenology in the stress response under a two--step protocol. Our model system also recovers the behaviour under a more recently studied three--step protocol, which consists of a compression followed by a decompression and then a final compression. We show that the observed response in both two--step and three--step protocols can be understood using Linear Response Theory. In particular, a linear scaling with age for the two-step protocol arises generically for slow linear responses with power law or logarithmic decay and does not in itself point to any underlying aging dynamics.
\end{abstract}

\maketitle

Memory in the context of physical systems refers to the ability to encode, access, and erase signatures of past history in the state of a system~\cite{keim2019memory}. This ability is absent when the system is in thermal equilibrium and consequently it is tied to out-of-equilibrium behavior. The origin and process of memory formation have been an intense area of research in last decade. In the context of jamming, the origin of memory in hard sphere glasses has been linked to a Gardner transition~\cite{morse21} very recently. Researchers have also explored memory formation in disordered media through directed aging~\cite{nagel19, nagel20} or simply through periodic driving~\cite{sastry14, nagel21}. In structural glasses, memory effects have also been studied~\cite{scalliet2019rejuvenation} through temperature cycles, which have helped to unveil the complex hierarchical structure of the free energy landscape; see Ref.~\cite{arceri20} for a discussion of recent advances in the study of memory and rejuvenation effects in structural glasses.

A remarkable experimental protocol introduced by Kovacs~\cite{kovacs1963glass} for the analysis of polymer glasses revealed that the time--dependent evolution of a glassy system can depend sensitively on its thermal history. A prime example is the non-monotonic evolution of a macroscopic observable (e.g.~volume) after rapidly cooling an initially equilibrated sample, allowing it to relax for some duration and then warming it up instantaneously to a higher temperature; this has since then been dubbed ``Kovacs effect'' and has been analyzed theoretically for different models of glasses~\cite{cugliandolo2004memory, bertin2003kovacs, activation20}, also within the framework of Linear Response Theory (LRT)~\cite{prados2010kovacs}. 

A generalization of the protocol described above has received increasing attention over the last few years, whereby the control parameter is generally different from temperature and the dynamics can be athermal. The systems that have been considered in this context are quite diverse, ranging from foams to crumpled metallic sheets and even jammed glass beads~\cite{lahini2017nonmonotonic, kursten2017giant, rubinstein2018nonmonotonic,he2019,prados20, murphy2020memory,omar20}. These systems typically show logarithmic stress relaxation when compressed. This slow relaxation can be understood as a collective effect coming from a broad distribution of relaxation rates of the system (see~\cite{amir2012relaxations} for details). The common effect observed in all of those systems is the non-monotonic relaxation of the relevant response variable associated with the control parameter. To be specific, in a two--step protocol, where a compression is held for a waiting time $t_{\rm w}$ and then the system is decompressed, a non-monotonic response of the pressure is observed after the second perturbation, i.e.\ for $t>t_{\rm{w}}$. It turns out that the time $t_p$ where the peak in the response occurs grows with the waiting time $t_{\rm w}$. In Refs.~\cite{lahini2017nonmonotonic,he2019} this scaling is reported as linear. In a recent study~\cite{murphy2020memory} a three--step protocol (which adds a final compression after the two previous step perturbations) was explored in athermal systems and found to produce a more complicated behaviour, with the response exhibiting two extrema (one maximum and one minimum) instead of one. Nonetheless, the position of the maximum again scaled with the first waiting time, therefore showing some similarity with the two--step protocol as regards memory effects. 

\begin{figure}
\includegraphics[height =0.45\linewidth]{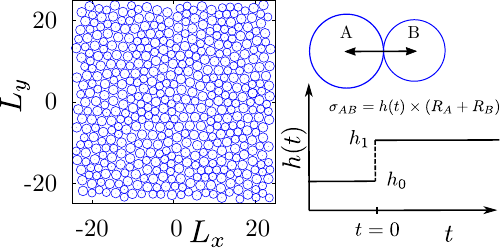}
\caption{(Left) Typical snapshot of a portion of the athermal binary mixture of harmonic particles. Compression and decompression are implemented by scaling the diameter of all particles by a time dependent scaling factor, $h(t)$ (sketch top right). (Right) Schematic of the control parameter $h(t)$ for the one--step protocol.} 
\label{schematic}
\end{figure} 
In spite of a substantial research effort aimed at understanding such memory effects and non-monotonic responses, the field still lacks a simple particle-based simulation model. In this paper we show that a well-known athermal model system~\cite{chacko19}  efficiently and accurately recovers the phenomenology of most relevant experimental scenarios, {\textit{i.e.}}\ one--step, two--step and three--step deformation protocols, using simple numerical simulations. To be specific we recover the logarithmic stress relaxation when the system is compressed and the non-monotonic stress response for the Kovacs-like two-step protocol. Our simulations also exhibit a linear scaling of  $t_p$ with $t_{\rm w}$ to leading order, as well as further aspects of the phenomenology recently observed under three--step protocols. We emphasise the usefulness of having such an {\textit{in-silico}} experimental system as the simulations are easily reproducible and one can in principle measure every dynamical and structural aspect, which could be very challenging or infeasible in an actual experiment with a real athermal solid like a foam or crumpled metallic sheet.

Remarkably, we are able to rationalize our results solely using Linear Response Theory, using an approach similar to~\cite{plata2017kovacs} but of course with the relevant response functions, which for step strain perturbations show an initial elastic response followed by a slow relaxation. An important theoretical insight will be that these response functions are {\em time translation-invariant}, which means that the linear memory effects we see here are, in spite of their superficial similarities, quite distinct from memory caused by an underlying aging dynamics as in glasses~\cite{kovacs1963glass, cugliandolo2004memory, bertin2003kovacs, activation20} We emphasize finally that we use the term LRT here in its broad sense of a linear relation between response and the corresponding perturbation. Such a linear response is generally expected for small perturbations, but this is not a consequence of the Fluctuation-Dissipation Theorem here: the latter is inapplicable for our athermal systems as there are no spontaneous fluctuations whose correlations could be measured.

This paper is structured as follows. In section~\ref{model_sim} we discuss the model (subsection~\ref{model}) and the different perturbation protocols (subsection~\ref{prots}). In section~\ref{sec:lin} we set up the Linear Response formalism. In section~\ref{sec:res} we show the results from simulations for each of the protocols (one--step, two--step and three--step) and we compare those with the predictions from LRT. Finally, in section~\ref{disc} we conclude with a summary and discussion.

\section{Model \& Simulation}
\label{model_sim}

\subsection{Model}
\label{model}

To model an athermal solid we use a binary mixture of soft particles interacting through a pairwise harmonic interaction (as used in Ref.~\cite{durianPRL1995,DurianPRE1997} and more recently in Ref.~\cite{chacko19}, see Fig.~\ref{schematic} for a typical snapshot). The particles interact only when they overlap; explicitly the interaction potential is
\begin{equation}
    V(r_{ij}) =  \frac{1}{2} k R^3 \left(1-\frac{r_{ij}}{\sigma_{ij}}\right)^2
                        \Theta \left(r_{ij} - \sigma_{ij}\right)
\end{equation}
where $\sigma_{ij}$ is an additive interaction diameter computed as $\sigma_{ij}=h(t)\times(R_i+R_j)$. Here $R_i$ and $R_j$ are the radii of particles $i$ and $j$, which can be either $R_A=R$ or $R_B=1.4R$ depending on whether the particle concerned is of A-type or B-type ({\it cf.}~Fig.~\ref{schematic}). We fix our units by choosing for simplicity the force constant $k=1$ and the radius scale $R=1$. The time-dependent control parameter $h(t)$ determines how compressed the system is and {$\Theta(\cdot)$} is the Heaviside function. We simulate $N=10^4$ particles in a two-dimensional periodic square box with side length $L=216$, with equal numbers $N_A=N_B=N/2$ of A- and B-particles. For the dynamics we assume a standard athermal overdamped equation of motion with implicit friction against a stationary solvent,
\begin{equation}
     \frac{\mathrm{d}\mathbf{r}_i}{\mathrm{d}t}
    = - \frac{1}{\zeta} \mathbf{\nabla}_{i}  \sum \limits_{j \ne i}
      V(r_{ij})
 \label{eq:dynam}
\end{equation}
where $\mathbf{r}_i$ is the position of particle $i$ and $\nabla_i = \partial/\partial \mathbf{r}_i$. We fix the drag coefficient $\zeta=1$ from now on. This, combined with the unit choices for $R$ and $k$, then also sets the time unit $\zeta/(kR)=1$. To solve the particle dynamics given by Eq.~\ref{eq:dynam} numerically  we use an Euler scheme with fixed time step $\Delta t=0.01$. The packing fraction in our system is defined as 
\begin{equation}
\phi=\frac{N_A \pi {(h(t) R_A)}^2+N_B \pi {(h(t) R_B)}^2}{L^2} \propto h^2(t)
\end{equation}
 All dynamical quantities are averaged over $128$ independent simulation runs unless explicitly stated otherwise, starting as in Ref.~\cite{chacko19} with randomly distributed positions of the particles in the simulation box. We first prepare the system at area fraction $\phi=1$ (corresponding to $h=1$) and run the dynamics for a long time ($t=10^6$) to guarantee that the system reaches an energy minimum, {\it i.e.}\ a fully relaxed state. These minimised configurations are then used as the initial conditions (with $t$ reset to zero) for all perturbation protocols considered below.

\begin{figure}
\centering
\includegraphics[height = 0.75\linewidth]{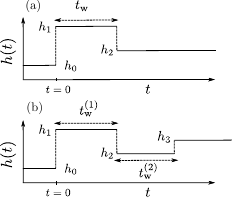}
\caption{(a) Two--step protocol. At time $t=0$ the control parameter $h$ is suddenly switched from $h_0$ to $h_1$ {and kept at that value} for a time {duration} $t_{\rm{w}}$. Then it is changed {again} to the final value $h_2$. (b) Three--step protocol. In addition to the steps in (a) (with $t_{\rm{w}} \to t_{\rm{w}}^{(1)}$), the value $h_2$ is maintained for a time $t_{\rm{w}}^{(2)}$ after which a {final change of $h(t)$ to $h_3$} is performed.}
\label{protocols}
\end{figure}

\subsection{Protocols}
\label{prots}

To explore the memory-like response we deform the system using three different protocols. These are described below and shown schematically in Figs.~\ref{schematic} and~\ref{protocols}. The protocols are named according to the number of times in which the control parameter {$h(t)$} is changed instantaneously.

\subsubsection{One--Step Protocol} 
In the \emph{one--step protocol} the system starts in the fully relaxed state associated with $h_0 = 1.0$ and then at time $t=0$ this parameter is suddenly changed to $h_1 > h_0$ (right panel of Fig.~\ref{schematic}). This causes a compression of the system: the  increase in $h$ increases the effective particle sizes and so leads to a higher area fraction as the box size remains the same. 

The conjugate variable to $h$ is the virial pressure, which is  the negative diagonal component of the virial stress tensor
 \begin{equation}
      P(t)=-(\Sigma_{xx}+\Sigma_{yy})/2 \,, \qquad \Sigma_{\alpha \beta}=-L^{-2}\sum^{N}_{i=1} r^\alpha_i F^\beta_i
      \label{pressure}
 \end{equation}
where $\mathbf{F}_i$ the net force on the $i$-th particle.
From symmetry $\langle \Sigma_{xx} \rangle=\langle \Sigma_{yy} \rangle$; numerically we therefore use only  $\Sigma_{xx}$ to evaluate $P(t)$. In our {simulations} the relaxation of this quantity is tracked as a function of time. This protocol provides the relaxation curves (see Fig.~\ref{fig:stressdecay}) that will be used for the Linear Response prediction, as discussed below in section~\ref{sec:lin}.

\subsubsection{Two--Step Protocol} 

In the \emph{two--step} protocol one starts again with the fully relaxed system at $h_0=1.0$. The control parameter $h(t)$ is switched at $t=0$ from $h_0$ to a higher value $h_1$, as in the one--step protocol. The system is then allowed to relax in that state for a waiting time $t_{\rm w}$. At this point one makes a further change by decompressing the system, reducing $h(t)$ from $h_1$ to a new value $h_2$ at $t=t_{\rm w}$; we consider $ h_0 < h_2 < h_1$ (see Fig.~\ref{protocols}(a) for a sketch). Similar to the one--step protocol we monitor the pressure $P(t)$ as an observable. The slow decay of $P(t)$ up to time $t_{\rm w}$, which is the same as in the one--step protocol, is ``interrupted'' by the sudden drop in area fraction at $t=t_{\rm w}$. We analyze the evolution of $P(t)$ after this point {through direct numerical simulation and also using LRT (see section~\ref{sec:res})}.

\subsubsection{Three--Step Protocol} Finally in the \emph{three--step protocol} one proceeds as in the two--step protocol, with the time during which $h(t)$ is held at $h_1$ now denoted $t_{\rm w}^{(1)}$. After the second step, $h(t)$ is left constant at $h_2$ for a further time interval $t_{\rm w}^{(2)}$ and at time $t=t_{\rm w}^{(1)} + t_{\rm w}^{(2)}$ the parameter $h(t)$ is  switched for a third time, from $h_2$ to $h_3$. We assume again that this value lies between the two previous ones, $h_0<h_2<h_3<h_1$ (see Fig.~\ref{protocols}(b) for a schematic). Hence the system remains at $h(t)=h_1$ for a time interval $t_{\rm w}^{(1)}$ and at $h(t)=h_2$ for a time interval $t_{\rm w}^{(2)}$. The pressure $P(t)$ is again monitored after the final perturbation through direct numerical simulation and compared with LRT (see section~\ref{sec:res}). 

In real experiments, particle sizes -- or equivalently system volume at fixed particle size -- cannot be changed instantaneously. Nonetheless these changes can often be performed rapidly enough so that internal relaxation processes can be neglected while they take place. The instantaneous changes of $h(t)$ that we work with constitute the idealized limit of such a scenario. 
\begin{figure}
\includegraphics[height = 0.65\linewidth]{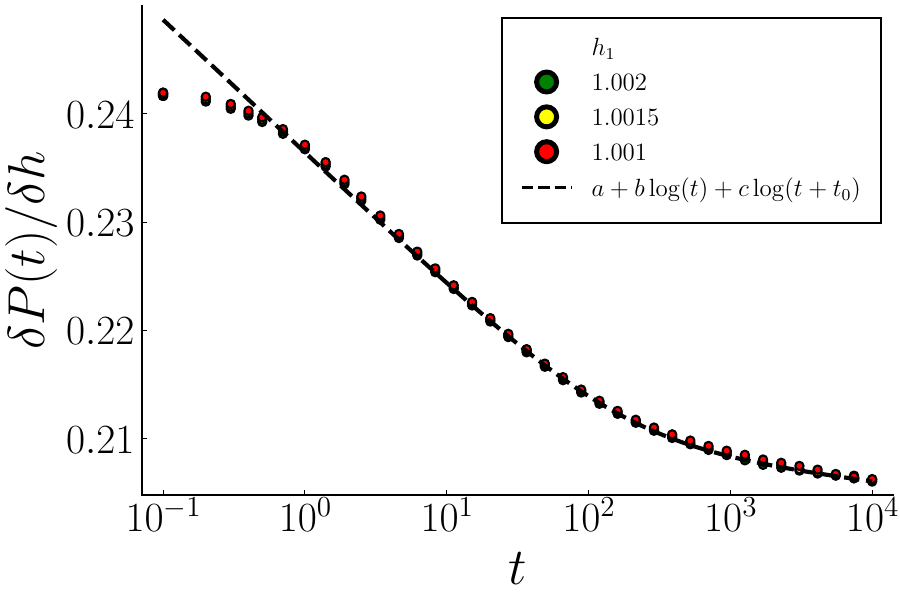}
\caption{Linear response regime. After complete relaxation at $h_0 = 1.0$ ($P(0) = 0.01563$) the scale factor $h$ is changed to a new value $h_1$ at time $t = 0$ following the one--step protocol (see Fig.~\ref{schematic}). The resulting response function
$\delta P(t)/\delta h$ (see  Eq.~\eqref{suscept})
is plotted as a function of time (markers: data, dashed line: double logarithmic fit (Eq.~\eqref{singlepfit})); results for different $h_1$ overlap almost completely, demonstrating linear response.  Fit parameters: $a = 0.2125, b = -0.00533, c= 0.00462, t_0 = 176.21$.}
\label{fig:stressdecay}
\end{figure}

\section{Linear Response Prediction}
\label{sec:lin}

LRT is a well grounded framework to analyze memory effects in thermal systems~\cite{bohmer1995pulsed, marconi2008fluctuation, prados2010kovacs,  diezemann2011memory}. In particular, Ref.~\cite{plata2017kovacs} illustrates the theory for Markovian athermal systems, as is the case here. For a general time-dependent perturbation one has the general linear response form~\cite{bedeaux1971linear, marconi2008fluctuation}
\begin{equation}
    \delta P(t) = \int_0^t dt'\,\chi(t-t')\dot h(t')
    \label{linres} 
\end{equation}
where $\delta P(t) = P(t)-P(0)$, $\dot h$ denotes the time derivative and $\chi(t)$ is called the step response function (also known as time-dependent susceptibility). This function is obtained from the one--step protocol and is used to predict the response for the multi--step protocols as we show shortly.

\emph{One--step protocol.}
For this protocol we set $h(t) = h_0 + \Theta(t) \delta h$ where $\delta h = h_1 - h_0$. Substitution of this in Eq.~\eqref{linres} retrieves
\begin{equation}
    \chi(t) = \frac{\delta P(t)}{\delta h}
    \label{suscept}
\end{equation}
In LRT this response function is independent of $\delta h$, the size of the perturbation.

\emph{Two--step protocol.} 
 For small changes $h_1 = h_0 + \delta h_1$ and $h_2 = h_1 - \delta h_2$ with $|\delta h_i/{h_i}| \ll 1$, it follows from Eq.~\eqref{linres} that
the pressure for $t > t_{\rm{w}}$ is given by the superposition of the relaxation function evaluated at different times, that is
\begin{equation}
     P(t) = P(0) + \delta h_1\, \chi(t) - \delta h_2\, \chi(t-t_{\rm{w}}) 
    \label{twopred}
\end{equation}

\emph{Three--step protocol.} In addition to the previous steps we consider here a further perturbation $h_3 = h_2 + \delta h_3$ with {$|{\delta h_3}/{h_3}| \ll 1$}.  Using again the linear superposition principle Eq.~\eqref{linres}, the pressure for $t > t_{\rm w}^{(1)} + t_{\rm w}^{(2)}$ becomes
\begin{align}
 P(t) &= P(0) + \delta h_1\, \chi(t) - {\delta h_2}\,\chi(t - t_{\rm w}^{(1)}) \notag \\
    &+ {\delta h_3}\,\chi(t - {t_{\rm w}^{(1)}} - t_{\rm w}^{(2)})  
     \label{threepred}
\end{align}
in terms of the same relaxation function $\chi(t)$ as before.
\begin{figure}  
\includegraphics[height =0.65\linewidth]{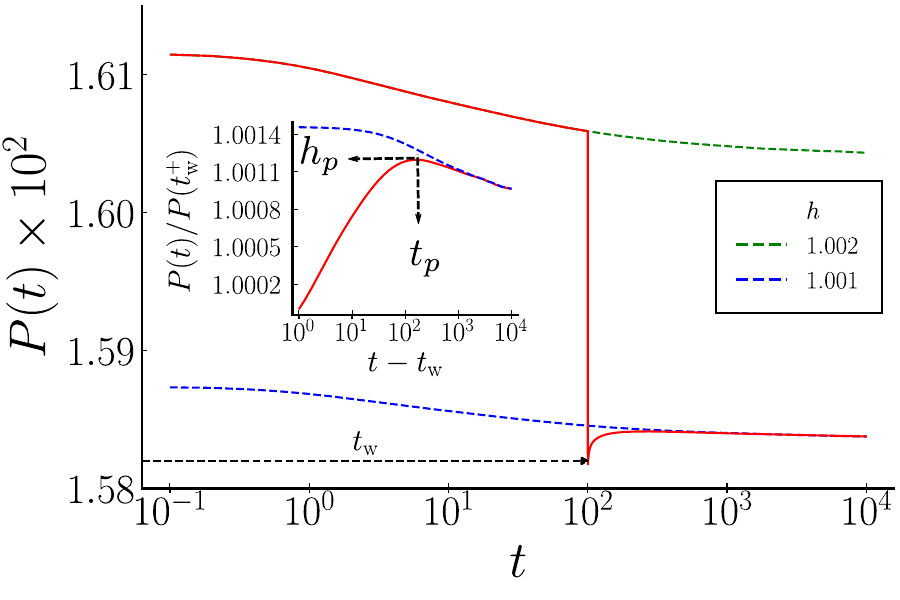}
\caption{Evolution of the pressure $P(t)$ for a two--step protocol with $t_{\rm w}=10^2$ and $h_1=1.002$, $h_2=1.001$. Dashed lines
  correspond to the evolution for the one--step protocol with the associated
values of $h$. Inset:  Non-monotonic response after the final (second) perturbation, showing pressure normalized by its value at $ t_{\rm{w}}^+ :=t_{\rm w} + 1$, {\it i.e.}\ just after the second perturbation; the time of {the} maximum {of the} response $t_p$ and its corresponding normalized height $h_p$ are indicated.} 
\label{twostepa}
\end{figure}

\section{Results}
\label{sec:res}
 
In this section we show that our simulation results recover all relevant observations for multi--step protocols (two and three--step) as seen in recent experiments~\cite{lahini2017nonmonotonic, kursten2017giant, rubinstein2018nonmonotonic,he2019,prados20, murphy2020memory,omar20}. We first demonstrate that our simulations lie in the Linear Response limit, and then we show that using LRT we can correctly predict the observed phenomenology via Eqns.~\eqref{twopred} and~\eqref{threepred}. 

In addition we make an asymptotic prediction for the scaling with $t_{\rm{w}}$ of the time of the maximum of the response $t_p$ and its corresponding normalized height $h_p$ for the two--step protocol (see Fig.~\ref{twostepa}) using LRT. For this purpose we first fit the step response from the one--step protocol to a double logarithm. This fit was already suggested in the analysis of experimental data ({see Ref}.~\cite{lahini2017nonmonotonic}). We further improve the fit by including the  effects of the eventual saturation of the response function. Finally, we show that according to our numerical results we do not need an extra prefactor `$C$' in front of $t_{\rm w}^{(1)}$ and $t_{\rm w}^{(2)}$  (in Eq.~\eqref{threepred}) to fit our response function in the framework of LRT.

\begin{figure*}
\includegraphics[scale=1.43]{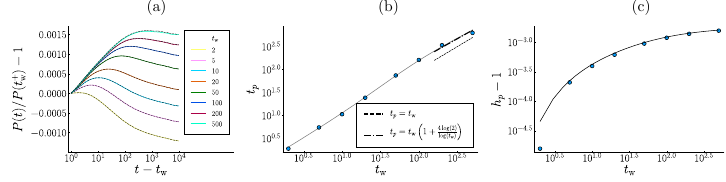}
\caption{Two--step protocol. $h_0 = 1.0$, $h_1 = 1.002$ and $h_2 = 1.001$ (a) Pressure as a function of time for different waiting times {from direct numerical simulations} (solid lines) together with linear response predictions (dashed lines). (b) Dependence of the maximum time $t_p$ on $t_{\rm w}$ (markers: data, solid line: linear response prediction, dashed line: leading order asymptotic prediction (Eq.~\eqref{asymptotictp}), dash--dotted line: asymptotic prediction with first order correction  (Eq.~\eqref{tpimpro}). (c) Dependence of {$h_p-1$} on $t_{\rm w}$ (markers: data, solid line: linear response prediction).} 
\label{twostepfig}
\end{figure*}  
 
\emph{One--step protocol.} After the perturbation at $t=0$ we observe a sudden pressure increase that comes from the elastic response of the amorphous solid, followed by a slow monotonic {decay of the pressure}. In Fig.~\ref{fig:stressdecay} we show that the observed response falls within the linear regime. This is established by the collapse of the ratio $\delta P (t)/\delta h$ for different values of $h_1$. This ratio corresponds to the step response (Eq.~\eqref{suscept}). We then fit it to  a double logarithmic curve of the form
\begin{align}
\chi(t) = a + b \log(t) + c\log(t+t_0)
\label{singlepfit}
\end{align}
This fit is correct in a restricted time window: Fig.~\ref{fig:stressdecay} shows that it becomes inaccurate at small times and Fig.~\ref{fits} demonstrates that it also fails for very long times; see also the discussion around Eq.~\eqref{tpimpro}. Note that as long as we can measure and evaluate the linear step response $\chi(t)$ in some way, such a fit is not required to be able to use Eq.~\eqref{twopred} or Eq.~\eqref{threepred}. However, a closed form expression for the step response allows analytical predictions as we will discuss shortly. 

\emph{Two--step protocol.} After the sudden drop at $t_{\rm w}$, again due to the elastic response to the step change in $h(t)$, the pressure $P(t)$ rises for some time and then decreases, eventually merging into the asymptotic curve from the one-step protocol for the same final $h_2$. This can be seen in Fig.~\ref{twostepa}, where the solid line shows the non-monotonic response for the two--step protocol while the dashed lines correspond to the relaxations one finds for the one--step protocol with final parameters $h_1$ and $h_2$ respectively. The location of the maximum of the two-step relaxation curve depends strongly on the age of the sample $t_{\rm w}$; intuitively, then, the system seems to ``remember'' how long it has been kept at the compression set by $h_1$ (see Fig.~\ref{twostepfig} for a quantitative comparison). Such memory effect has been observed in experiments on various athermal systems~\cite{lahini2017nonmonotonic, kursten2017giant, rubinstein2018nonmonotonic, murphy2020memory}.

We can use Eq.~\eqref{twopred} and the data for $\chi(t)$ derived from the simulation of the one--step protocol via Eq.~\eqref{suscept} to predict this nonmonotonic response (dashed lines in Fig.~\ref{twostepfig}(a)) {as a function} of $t_{\rm w}$. As the figure demonstrates, the prediction from LRT is very accurate, clearly supporting the  validity of LRT for our simulation setup. We remark that this prediction is made directly from the measured one-step response and does not require e.g.\ a parametric fit of this function. Such a fit is necessary, however, to predict analytically the scaling of $t_p$ and $h_p$ with $t_{\rm w}$ (solid lines in Fig.~\ref{twostepfig}(b) and Fig.~\ref{twostepfig}(c)). Here we define $t_p$ as the length of time between $t_{\rm w}$ and the maximum in the response. To find this, one  solves for the maximum of Eq.~\eqref{twopred} and subtracts $t_{\rm w}$. For the double logarithmic fit in Eq.~(\ref{singlepfit}) this yields a complicated but closed form expression that can be inserted back into Eq.~\eqref{twopred}. 

Following~\cite{lahini2017nonmonotonic} we define the height of the maximum $h_p$ by normalizing the result by $P(t_{\rm w}^+)$, the value of the pressure just after the time $t_{\rm w}$ of the second perturbation. Explicitly we use $P(t_{\rm{w}} + 1.0)$; the unit time difference is chosen here because it is the smallest time for which the logarithmic fit works well (see Fig.~\ref{fig:stressdecay}). The resulting expressions for $t_p$ and $h_p$ are rather long, but for  large  $t_{\rm w}$ (compared to the fit parameter $t_0$, which is of order $10^2$) one recovers a single logarithmic step response function (from Eq.~\eqref{singlepfit}) for which the scaling can easily be found.  In this case the solution for the position of the maximum of Eq.~\eqref{twopred} is given by 
\begin{align}
    t_p = t_{\rm w} \frac{\delta h_2}{{\delta h_1} - {\delta h_2} }
    \label{asymptotictp}
\end{align}
This relation was derived in Ref.~\cite{lahini2017nonmonotonic} from another perspective and is recovered here within the LRT.  Two main points can be made from the result in Eq.~\eqref{asymptotictp}: (i) the asymptotic prediction for the scaling of $t_p$ is linear in $t_{\rm{w}}$, (ii) the scaling factor is independent of any fitting parameters of the step response function and only depends on the relative strength of the two perturbations. For our data the relative ratio $\frac{\delta h_2}{{\delta h_1} - {\delta h_2} } = 1$ and it predicts that $t_p = t_{\rm w}$ (see dashed line in Fig.~\ref{twostepfig}(b)). As can be seen from Fig.~\ref{twostepfig}(b), however, corrections to this leading order asymptotic prediction remain visible within the time window considered in the simulations. 

The asymptotic prediction for the scaling of $t_p$ with $t_{\rm w}$ can be improved by considering a more realistic fit for the step response $\chi(t)$, which incorporates the physical expectation that the step response function should approach a constant (which can be zero) at long times. The simple fit in Eq.~(\ref{singlepfit}) does not have this property in the generic case $b\neq -c$. We choose then the modified fit function  $\chi(t) = \tilde{a} + \frac{\tilde{b} \log(t)}{1+\tilde{c}\log(t+ \tilde{t}_0)}$. In Appendix~\ref{ap:fit} we show that this function provides an adequate fit of the data for the step response for intermediate and long times (Fig.~\ref{fits}). The asymptotic prediction based on this fit (worked out in Appendix~\ref{sec:generic}) is 
\begin{align}
    t_p = t_{\rm w} \frac{\delta h_2}{{\delta h_1} - {\delta h_2} } \left( 1 +  \frac{2 \delta h_1}{{\delta h_1} - {\delta h_2}} \frac{\log (\delta h_1/\delta h_2) }{\log(t_{\rm w})}  \right)
    \label{tpimpro}
\end{align}
This agrees with the leading order term in Eq.~(\ref{asymptotictp}) but also contains a relative correction of order $1/\log(t_{\rm w})$.
For our chosen set of parameters it quantitatively evaluates to the dash-dotted line in Figure~\ref{twostepfig}(b) and gives a significantly improved prediction for $t_p$ that matches the numerical data in the long waiting time regime. 
\begin{figure*}
\includegraphics[scale=1.43]{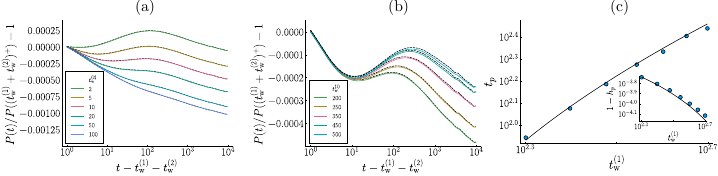}
	\caption{Three--step protocol.  $h_0 = 1.0$, $h_1 = 1.002$, $h_2 = 1.001$, $h_3 = 1.0015$. Shown are averages over a set of $512$ independent trajectories. (a) Fixed first waiting time $t_{\rm w}^{(1)} = 200$, second waiting time $t_{\rm w}^{(2)}$ as shown in legend. (b) Fixed second waiting time $t_{\rm w}^{(2)} = 10$, first waiting time $t_{\rm w}^{(1)}$ as shown in legend. For (a) and (b), {solid lines: direct numerical simulations}, dashed lines: linear response prediction. The pressure has been normalized by its value just after the last step perturbation, {\it i.e.}\ at $(t_{\rm w}^{(1)} + t_{\rm w}^{(2)})^+:=t_{\rm w}^{(1)} + t_{\rm w}^{(2)}+1.0$. (c) $t_{\rm w}$-scaling of $t_p$ and $h_p$ for the data in (b) (markers: simulation data, solid line: linear response prediction). The correlations between the small statistical  fluctuations appearing at very long times $\sim 10^3-10^4$ in subfigures (a) and (b) arise from the fact that the same set of initial configurations was used for the simulations with different $t_{\rm w}^{(1)}$ and $t_{\rm w}^{(2)}$.}
\label{threestepfig}
\end{figure*} 

\emph{Three--step protocol.} 
In the three--step protocol we have two control parameters for the timing of the perturbations; $t_{\rm w}^{(1)}$ and $t_{\rm w}^{(2)}$. Depending on the combination of these waiting times a non-monotonic evolution of the pressure $P(t)$ can again be observed (see Fig.~\ref{threestepfig}(a),(b)).

In the first scenario we keep $t_{\rm w}^{(1)}$ constant and vary $t_{\rm w}^{(2)}$. The non-monotonicity of the response is evident for small values of  $t_{\rm w}^{(2)}$; this then crosses over to a monotonic decay for long values of the second waiting time (see Fig.~\ref{threestepfig}(a)). Still the LRT works perfectly well (dashed lines in Fig.~\ref{threestepfig}(a)). The linear response prediction is obtained by again using the response measured for a single step, together with Eq.~\eqref{threepred}.

In a second scenario we keep $t_{\rm w}^{(2)}$  constant and vary $t_{\rm w}^{(1)}$ instead. Here we observe two extrema in the response after the final perturbation, a minimum followed by a maximum. The maximum shifts to larger time differences as we increase the first waiting time $t^{(1)}_{\rm w}$, which is reminiscent of the behavior we saw for the two--step protocol. The position of the minimum, on the other hand, remains constant (see Fig.~\ref{threestepfig}(b)). Again, the linear response prediction works well without any fitting (dashed lines in Fig.~\ref{threestepfig}(b)). In this case analytical {closed form} expressions cannot be obtained for the dependence of $t_p$ and $h_p$ on $t_{\rm w}^{(1)}$, not even for asymptotically large $t_{\rm w}^{(1)}$. Nonetheless we can of course solve numerically for the maximum of Eq.~\eqref{threepred} to arrive at the predictions shown in Fig.~\ref{threestepfig}(c). The qualitative behaviour of $t_p$ and $h_p$ is the same as in the two--step protocol: both increase with the waiting time $t_{\rm w}^{(1) }$, thus the system also exhibits memory effects under this three--step protocol. 

\section{Discussion}
\label{disc}

In this work we demonstrate that a model athermal solid made of a binary mixture of soft (harmonic) particles can serve as a canonical model for understanding the non-monotonic response seen recently in a {diverse} type of athermal systems subjected to multi--step variations of the control parameter. Although the particular model we study has been used before to understand the rheology~\cite{durianPRL1995, DurianPRE1997} and aging of athermal systems~\cite{chacko19}, we have deployed it here for the first time to explore and understand two--step and three--step non-equilibrium experiments similar in spirit to the  Kovacs protocol.

Our results show that this athermal system exhibits memory-like effects that can be understood from Linear Response Theory. Indeed, as long as we are in the regime of small perturbations, the only ingredient needed for the LRT is the step response function (Eq.~\eqref{suscept}). If {the decay of this function is} slow enough (compared to e.g.\ an exponential) a non-monotonic response is expected for a two--step or three--step protocol. This follows from the structure of Eq.~\eqref{twopred} and Eq.~\eqref{threepred} (the same observation was also pointed out in Ref.~\cite{diezemann2011memory}). We note that we have checked the generality of this statement by considering a particle model with a different, Hertzian interaction. We found that it also exhibits a logarithmic decay of the pressure response to compression, and would therefore expect qualitatively the same phenomenology under multi--step protocols as for the harmonic repulsion considered above.

For the Kovacs--like two--step protocol, the way in which memory is displayed is via a relation between $t_{\rm w}$ and $t_p$ (or $h_p$). 
Using for the susceptibility the logarithmic fit function in Eq.~\eqref{singlepfit}, as proposed previously for experimental data in Ref.~\cite{lahini2017nonmonotonic}, an asymptotic relation for large $t_{\rm w}$ can be derived (Eq.~\eqref{asymptotictp}). This relation is linear and independent of the parameters used in the logarithmic fit of the step response function $\chi(t)$. This linear scaling has already been observed in athermal systems and its origin lies in the logarithmic behavior of the 
step response function (see for instance Refs.~\cite{lahini2017nonmonotonic, he2019}). Here we have derived this scaling on the basis of Linear Response Theory, which constitutes a new result in itself.

Two comments are in order as regards the above observations.  Firstly, one may wonder how generic the linear scaling of $t_p$ with $t_{\rm w}$ is. We show in appendix~\ref{sec:generic} that this linear scaling applies surprisingly broadly, i.e.\ whenever the step response function decays as a power law or a power of a logarithm. Secondly, we expect that for long waiting times the system actually reaches a steady state so that the response will reach a limiting value; a double logarithmic fit as in Eq.~(\ref{singlepfit}) therefore has to be an approximation that applies only for a finite time window, e.g.\ the one that is observable in the simulations. On this basis we proposed a fit function that takes into account the saturation of the response; for this a subleading correction to the linear scaling can then be derived, see Eq.~\eqref{tpimpro}. We showed that this modified fit function does an appreciably better job in terms of prediction of our simulation data (see Fig.~\ref{twostepfig}(b)).

In previous studies, the non-monotonic response observed in the two--step protocol was explained via the assumption of a broad distribution of relaxation times~\cite{lahini2017nonmonotonic, murphy2020memory}. This provides intuition for the initially unexpected non-monotonicity: consider a system (as for instance a glass~\cite{amir2012relaxations}) that is composed of slow and fast elements (with respect to experimental timescales), and that is subjected to the two--step protocol. Directly after the second perturbation, the system remembers -- via the slow elements -- its state before the perturbation, while the fast elements adapt rapidly to the new value of the external parameter. This creates a situation in which the two types of elements relax in opposite directions. The fast elements relax first, causing the unexpected increase in the response (in our case: pressure) with time. Eventually the slow elements then turn the relaxation around, giving rise to non-monotonic dynamics. This picture, though physically appealing, is arguably not needed to deduce the explicit formula in Eq.~\eqref{twopred}: the only ingredient required here is linear response, and we would argue that the earlier approaches were effectively using LRT, at least implicitly. To be specific, Eq.~(3) from Ref.~\cite{lahini2017nonmonotonic} can be recast in terms of the Linear Response formula (Eq.~\eqref{twopred}) with the susceptibility a single logarithmic decay function. On the other hand, it is easy to check that LRT with an exponential relaxation function {\em cannot} produce a non-monotonic response in a two-step protocol; what is needed is a slow relaxation, e.g.\ of power law or logarithmic type. In this sense the LRT framework helps to sharpen the physical picture of slow and fast elements into quantitative conditions on the relaxation function.

Overall, the memory effect we observe, which has also been reported for a range of different of systems~\cite{lahini2017nonmonotonic, kursten2017giant, rubinstein2018nonmonotonic,he2019,prados20, murphy2020memory,omar20}, can be explained using linear response as long as the perturbations are small. A natural question is what happens beyond the linear regime. As shown in Appendix~\ref{nonlinear}, qualitatively the same phenomenology is present in the non--linear regime. This indicates that the memory effect is a generic response of the system under the chosen protocols and not merely an artifact of linear response. A quantitative analysis of the $t_{\rm{p}}$ scaling in this regime is left for future work. 

Finally, it is worth discussing to what extent the physics of non-monotonic response to two-step or three-step protocols is related to memory and aging. As mentioned above, a linear response picture with a slow response function can be interpreted in terms of the dynamics of slow and fast modes: the slow modes can then reasonably be said to provide `memory' of the history of the system, even though this is in the form of a weak (linear) perturbation of the system. In the non--linear regime discussed above, the different perturbations can cause
irreversible structural changes, which are even more intuitive carriers of memory (see Fig.~\ref{irreversible} and Appendix~\ref{nonlinear} for more details).

An interpretation in terms of aging may also seem tempting, given that the maximum position $t_p$ grows with the age $t_{\rm w}$ of the system, in fact with a linear scaling that is often referred to as simple aging~\cite{monthus1996models, da2015slow}. However, given that our results are very well represented by a linear response theory with a 
response function that is {\em time-translation invariant}, {\it i.e.}\ depends only on time differences, the dynamics does not meet the conventional definition of aging, where correlations and response become genuine two-time functions~\cite{leticia97}. In this context it is interesting to note that in recent experimental studies ~\cite{lahini2017nonmonotonic, murphy2020memory} a constant factor $C$ had to be introduced \textit{ad hoc} in front of $t_{\rm w}$ in Eq.~\eqref{twopred} or in front of $t_{\rm w}^{(1)}$ and $t_{\rm w}^{(2)}$ in Eq.~\eqref{threepred} to fit the results of two-step and three-step protocols. This could indicate genuine aging behaviour, or alternatively effects that go beyond the linear response regime. 

We believe that in order to arrive at a full understanding of the type of memory analyzed in this paper for athermal systems, the gap between experiments and theory needs to be closed by simulation models that allow detailed inspection of the microscopic dynamics. {We hope that our work paves the way} for future work in this direction. An obvious avenue in that direction would be to extend our current study to initial conditions that are not yet arrested and lie in the athermal aging regime \cite{chacko19}, to understand the extent and impact of true aging effects on memory and non-monotonic response.

\paragraph*{Acknowledgement:}

We thank Jack Parley and Tunrayo Adeleke Larodo for useful discussions and comments on the manuscript. This project has received funding from the European Union’s Horizon 2020 research and innovation programme under the Marie Sk\l odowska-Curie grant agreement No 893128.

\section{Appendix}

\subsection{Fit of simulation data}
\label{ap:fit}
\begin{figure}  
\includegraphics[height =0.65\linewidth]{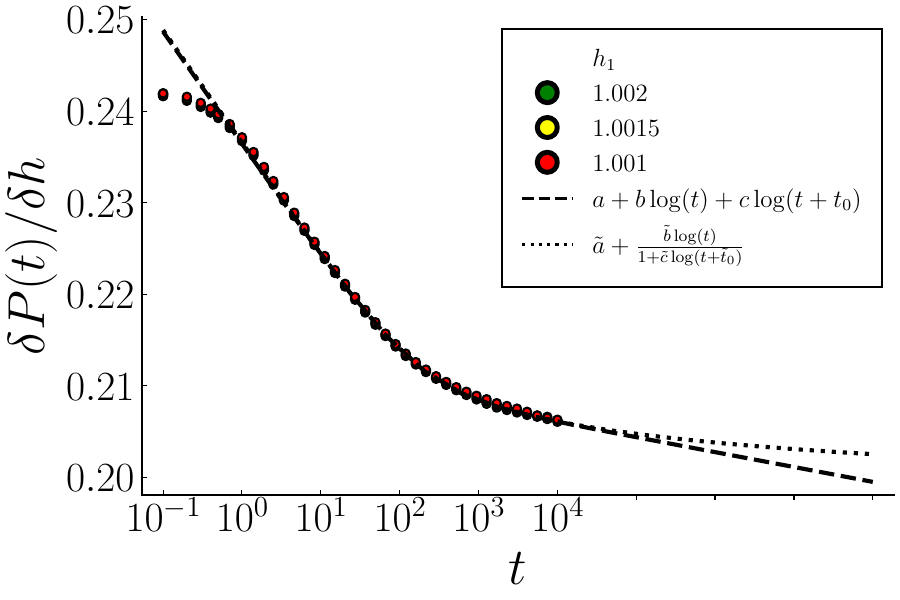}
\caption{Extension of Fig.~\ref{fig:stressdecay}, where now the fit provided by Eq.~\ref{extendedfit} has been added (dotted line).  In addition, the $x$-axis has been extended to show the fit prediction in the large time limit. Fit parameters: $\tilde{a} =  0.23658$,
$\tilde{b} = -0.01593$, $\tilde{c} = 0.4137, \tilde{t}_0 = 122.67$}
\label{fits}
\end{figure}

In the derivation of Eq.~\eqref{tpimpro} we introduced a modified fit function for the susceptibility, namely 
\begin{align}
    \chi(t) = \tilde{a} + \frac{\tilde{b} \log(t)}{1+\tilde{c} \log(t+ \tilde{t}_0)}
    \label{extendedfit}
\end{align}
In Fig.~\ref{fits} we show the fit of the simulation data to this functional form and its extrapolation to long times to compared with the double logarithm (Eq.~\eqref{singlepfit}).

For long times $t \gg \tilde{t}_0$, one may approximate Eq.~\eqref{extendedfit} as
\begin{align}
     \chi(t) &\approx \tilde{a} + \frac{\tilde{b}}{\tilde{c}} \left(\frac{1}{1 + \frac{1}{\tilde{c} \log(t)}} \right) \notag \\
     &= \tilde{a} + \frac{\tilde{b}}{\tilde{c}} \left( 1 - \frac{1}{\tilde{c} \log(t)} \right) + O\left(\frac{1}{\log^2(t)}\right) 
     \label{invlog}
\end{align}
From this expression one easily reads off that the step response saturates at the value $\chi(t \to \infty) = \tilde{a} + {\tilde{b}}/{\tilde{c}}$. This result is physically sensible: in contrast to a liquid (where the susceptibility decays to zero in the long time limit), an athermal solid can only relax part of the stress generated by a perturbation.

For short times, both expressions (Eq.~\eqref{singlepfit} and Eq.~\eqref{extendedfit}) are inconsistent with the data: as the material has a short time elastic response with a finite bulk modulus, $\chi(t=0)$ should be finite. We ignore this point as it has no impact on the predictions for multi--step protocols that we discuss in the main text.

\subsection{Generic linear relation between $t_p$ and $t_{\rm w}$ for some non--exponential response functions}
\label{sec:generic}

In this section we consider different non--exponential relaxation (step response) functions and obtain the associated scaling of $t_p$ with $t_{\rm{w}}$ in the large $t_{\rm{w}}$ limit. 

We start by considering a simple power law of the form $\chi(t) = a + b/t$. Substitution of this in the response for the two--step protocol (Eq.~\eqref{twopred}) yields
\begin{align}
  P(t) =  \delta h_1 \left(a + \frac{b}{t }  \right)  - \delta h_2 
   \left(a + \frac{b}{t  - t_{\rm{w}}}  \right)
\end{align}
The position $t^*$ of the maximum of this function ($P'(t^*) = 0$) is easy to determine; we subtract $t_{\rm{w}}$ to arrive at 
\begin{align}
t_p &= c_0 t_{\rm w} \label{linear} \\
&= \frac{t_{\rm{w}}}{\sqrt{ r} - 1}
\label{simplepower}
\end{align}
with $r := {\delta h_1}/{\delta h_2}$.

We can repeat these steps for a generic power law of the form $\chi(t) = a + bt^{-\alpha}$ with $\alpha > 0$. Then one arrives at the same linear form as in Eq.~\eqref{linear} but with a modified prefactor 
\begin{align}
c_0 = \frac{1}{ r^{1/(1+\alpha)} - 1}
\label{copower}
\end{align}
For $\alpha = 1$ we recover Eq.~\eqref{simplepower} as expected.

Another realistic non--exponential response functions is an inverse logarithm of the form  $\chi(t) = A + {B}/{\log(t)}$. This is the asymptotic limit of Eq.~\eqref{extendedfit} as derived in Eq.~\eqref{invlog} and yields the formula given in Eq.~\eqref{tpimpro} in the manuscript as we are going to show.

In this case one cannot solve explicitly for the maximum of the two--step response $P(t)$ (Eq.~\eqref{twopred}); instead the following implicit condition for $t_p$ is obtained:
\begin{align}
    \frac{\delta h_2}{t_p (\log(t_p))^2} = \frac{\delta h_1}{(t_p + t_{\rm{w}})( \log(t_p + t_{\rm{w}}))^2}
    \label{implicit}
\end{align} 
By extending the simple expression from Eq.~\eqref{linear} one can use as an \emph{ansatz} for $t_p$  a linear term plus a correction term as the inverse of the logarithm of the age:
\begin{align}
    t_p = t_{\rm w} \left(c_0 + \frac{c_1}{\log(t_{\rm{w}})} \right)
    \label{ansatz}
\end{align}
and then one has to solve for the unknowns $c_0$ and $c_1$. Substitution of Eq.~\eqref{ansatz} into expression given in Eq.~\eqref{implicit} leads to
\begin{align}
    \frac{(1+ c_0 + c_1/\log t_{\rm{w}}) (\log(t_{\rm{w}} (1+ c_0 + c_1/\log t_{\rm{w}}) ))^2 }{r (c_0 + c_1/\log t_{\rm{w}})  (\log( t_{\rm{w}} ( c_0 + c_1/\log t_{\rm{w}})) )^2} = 1
\end{align} 
It is then convenient to perform the change of variable $t_{\rm w} = \exp(l)$:
\begin{align}
    \frac{  (l+ c_0 l + c_1) (l + \log(1+ c_0 + c_1/l))^2 }{r  (c_0 l  + c_1) (l + \log(c_0  + c_1/l) )^2} = 1
\end{align} 
Since we are interested in an asymptotic relation (for large $t_{\rm{w}}$) we now perform a series expansion in $1/l$, which yields up to first order
\begin{align}
 \frac{1}{r} \bigg( \frac{(1+ c_0)}{c_0}  &-  \frac{c_1 + 2 c_0 (1 + c_0)\log(c_0)}{c_0^2 l}  \notag \\
 + &\frac{2 c_0 (1 + c_0) \log(1 + c_0)}{c_0^2 l}\bigg)  = 1
\end{align}
Solving now for the unknowns $c_0$ and $c_1$, we get first from the zeroth order term
\begin{align}
    c_0 = \frac{1}{r - 1}
    \label{c0}
\end{align}
Notice that this is consistent with Eq.~\eqref{copower} for $\alpha = 0$. From the first order term (in $1/l$) the coefficient of the leading correction term becomes
\begin{align}
    c_1 = 2 \frac{r}{(r - 1)^2 } \log \left( r \right)
    \label{c1}
\end{align}
Putting Eqs.~\eqref{c0} and~\eqref{c1} back into Eq.~\eqref{ansatz} one then obtains Eq.~\eqref{tpimpro} given in the main text.

As a further step we note that the previous result can be generalized to step response functions $\chi(t) = A + B (\log (t))^{-\beta}$ with $\beta >0$. Following the previous steps with the same ansatz (Eq.~\eqref{ansatz}) one arrives at the following expressions for $c_0$ and $c_1$:
\begin{align}
     c_0 = \frac{1}{r - 1} \, , \quad c_1 = 
     (1+\beta) \frac{r}{(r - 1)^2 } \log \left( r \right)
\end{align}
The linear term is the same as for the simple inverse logarithm (Eq.~\eqref{c0}), whereas the subleading correction differs in the prefactor $1+\beta$; for $\beta=1$ the results are of course consistent with Eq.~\eqref{c1} for $\beta = 1$.  

Finally one can generalize the above arguments to relaxation functions that decay as power laws with logarithmic corrections, $\chi(t) = a + b t^{-\alpha}(\log t)^{-\beta}$. We omit the details and only note that in this case the ansatz of Eq.~\eqref{ansatz} still works, with $c_0$ given by Eq.~\eqref{copower} and $c_1$ a (rather complicated) function of $\alpha$ and $\beta$.

\subsection{Non-linear Response}
\label{nonlinear}

The non--linear regime is detected by varying the strength of the perturbation $\delta h = h_1- h_0$ in the one--step protocol. In Fig.~\ref{linres} we show the response $\delta P /\delta h$ for different $h_1$. The results suggest that the system enters the non--linear regime around $\delta h=0.002$ where deviations from the linear response limit become clearly visible.

\begin{figure}[h]
\includegraphics[width =.9\linewidth]{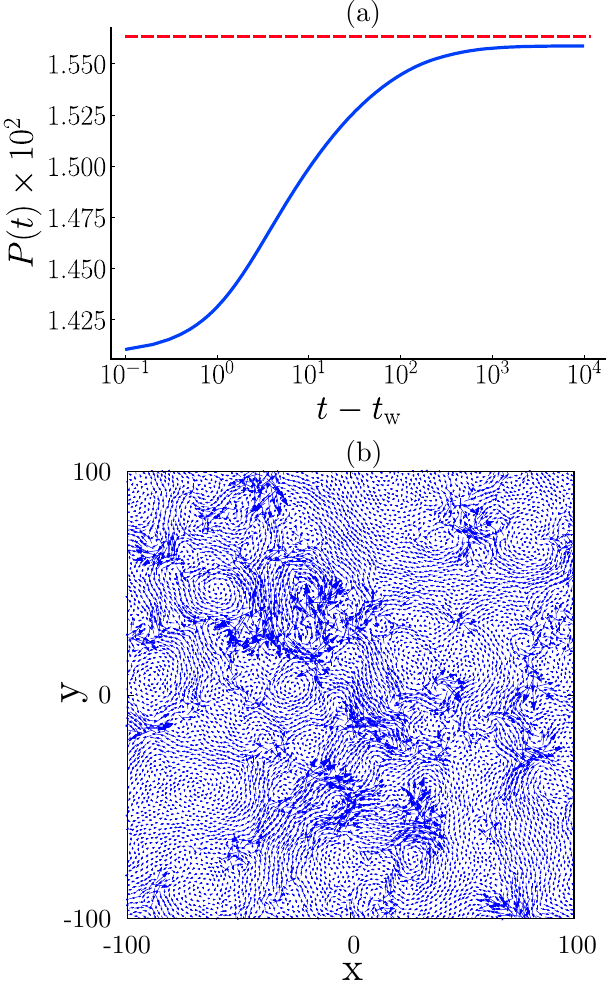}
\caption{(a) Pressure $P(t)$ (in blue solid line) of the system after a special two--step protocol with large deformation $h_0=1.0$, $h_1=1.1$, $h_2=1.0$ and $t_{\rm w}=2000$; with these parameters the protocol brings the perturbation parameter back to its initial state. Nonetheless the pressure does not return to its initial value (red dashed line) before the perturbation. (b) Displacement field of particles (scaled by a factor of $10$ for better visualisation) between the first perturbation and the last one clearly shows the irreversible changes in the particle configuration.} 
\label{irreversible}
\end{figure}

We first explored a special two--step protocol where $h_2=h_0$ and $h_1=1.1$. In this case one finds that even though the perturbation parameter is brought back to the original value, the transient perturbation causes {\em irreversible} changes in pressure (see Fig.~\ref{irreversible} (a)); the perturbation also causes nonzero particle displacements with a strongly non-affine pattern (see Fig.~\ref{irreversible} (b)), a clear indication of irreversible changes in the configuration.

\begin{figure}[h]
\includegraphics[width =\linewidth]{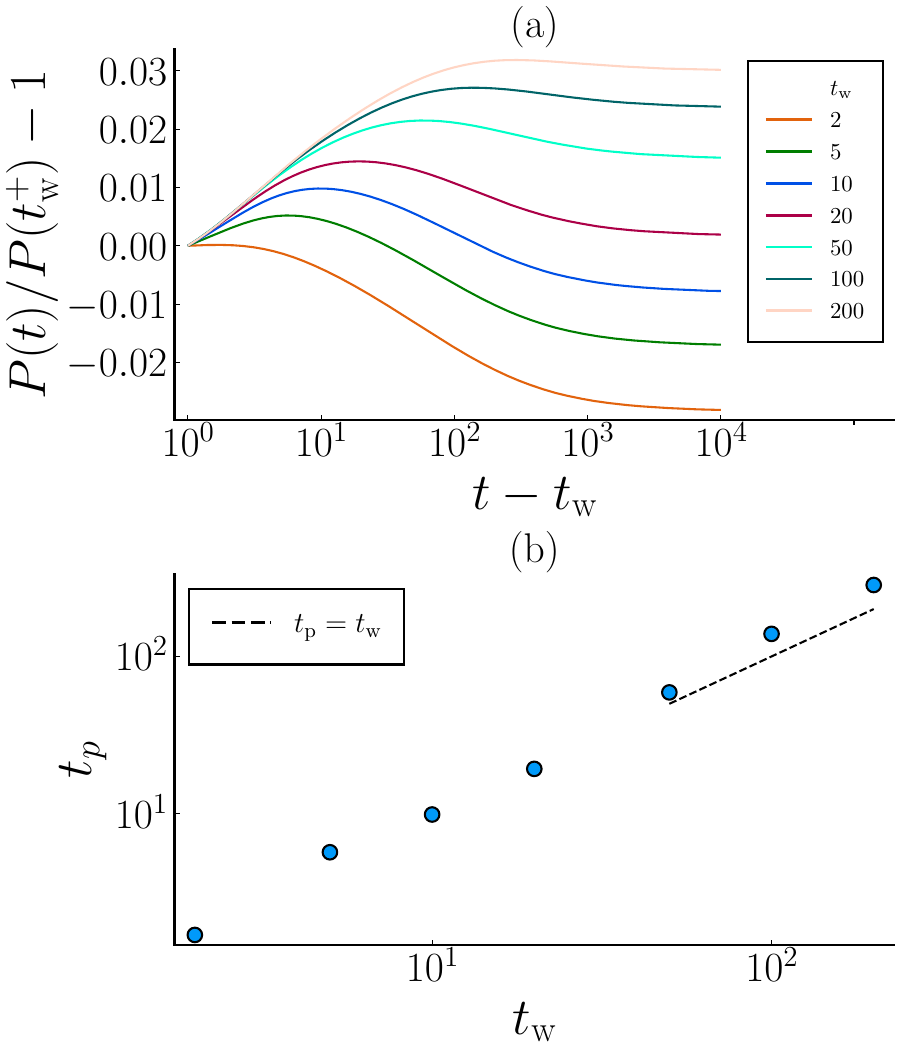}
\caption{Two--step protocol in the non-linear regime. $h_0 = 1.0$, $h_1 = 1.1$ and $h_2 = 1.05$ (a) Pressure as a function of time for different waiting times {from direct numerical simulations} (solid lines). (b) Dependence of the maximum time $t_p$ on $t_{\rm w}$ (markers: data).} 
\label{nonlintwo}
\end{figure}

To explore the memory phenomenology in the nonlinear regime, we have carried out the two--step protocol for different waiting times ($t_{\rm w}$). The results show a similar scenario as in the linear regime, both in terms of non-monotonic response (stress hump) and memory (linear dependence of $t_p$ on $t_{\rm w}$); see Fig.~\ref{nonlintwo}. This also points towards the fact that the memory effect we observe is quite generic and not restricted to the range of validity of  linear response theory.

\begin{figure}[h]
\includegraphics[height =0.55\linewidth]{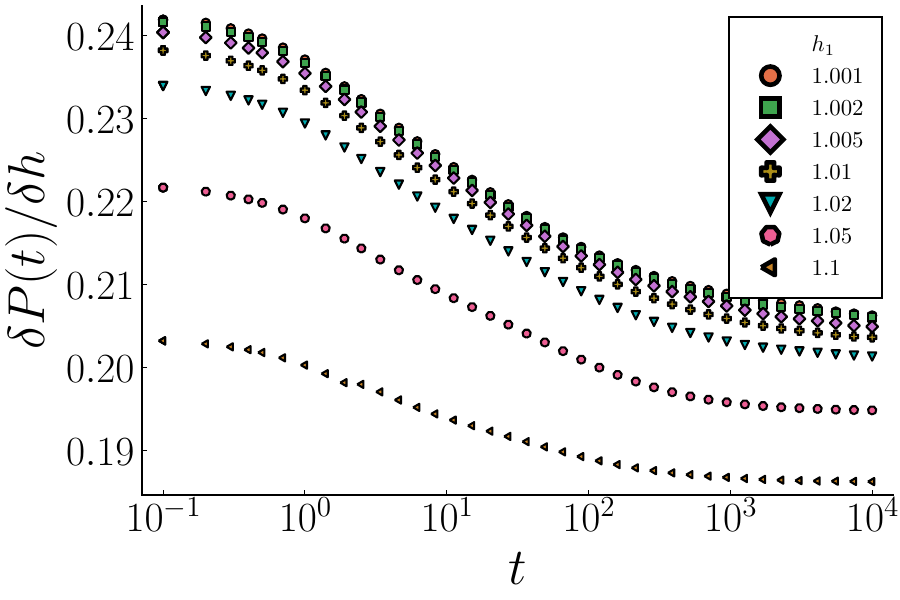}
\caption{Response function $\delta P(t)/\delta h$ for different values of $h_1$. Same initial condition as in Fig.~\ref{fig:stressdecay}.}
\label{linres}
\end{figure}

\bibliography{athermal}

\begin{thebibliography}{32}%
\makeatletter
\providecommand \@ifxundefined [1]{%
 \@ifx{#1\undefined}
}%
\providecommand \@ifnum [1]{%
 \ifnum #1\expandafter \@firstoftwo
 \else \expandafter \@secondoftwo
 \fi
}%
\providecommand \@ifx [1]{%
 \ifx #1\expandafter \@firstoftwo
 \else \expandafter \@secondoftwo
 \fi
}%
\providecommand \natexlab [1]{#1}%
\providecommand \enquote  [1]{``#1''}%
\providecommand \bibnamefont  [1]{#1}%
\providecommand \bibfnamefont [1]{#1}%
\providecommand \citenamefont [1]{#1}%
\providecommand \href@noop [0]{\@secondoftwo}%
\providecommand \href [0]{\begingroup \@sanitize@url \@href}%
\providecommand \@href[1]{\@@startlink{#1}\@@href}%
\providecommand \@@href[1]{\endgroup#1\@@endlink}%
\providecommand \@sanitize@url [0]{\catcode `\\12\catcode `\$12\catcode
  `\&12\catcode `\#12\catcode `\^12\catcode `\_12\catcode `\%12\relax}%
\providecommand \@@startlink[1]{}%
\providecommand \@@endlink[0]{}%
\providecommand \url  [0]{\begingroup\@sanitize@url \@url }%
\providecommand \@url [1]{\endgroup\@href {#1}{\urlprefix }}%
\providecommand \urlprefix  [0]{URL }%
\providecommand \Eprint [0]{\href }%
\providecommand \doibase [0]{http://dx.doi.org/}%
\providecommand \selectlanguage [0]{\@gobble}%
\providecommand \bibinfo  [0]{\@secondoftwo}%
\providecommand \bibfield  [0]{\@secondoftwo}%
\providecommand \translation [1]{[#1]}%
\providecommand \BibitemOpen [0]{}%
\providecommand \bibitemStop [0]{}%
\providecommand \bibitemNoStop [0]{.\EOS\space}%
\providecommand \EOS [0]{\spacefactor3000\relax}%
\providecommand \BibitemShut  [1]{\csname bibitem#1\endcsname}%
\let\auto@bib@innerbib\@empty
\bibitem [{\citenamefont {Keim}\ \emph {et~al.}(2019)\citenamefont {Keim},
  \citenamefont {Paulsen}, \citenamefont {Zeravcic}, \citenamefont {Sastry},\
  and\ \citenamefont {Nagel}}]{keim2019memory}%
  \BibitemOpen
  \bibfield  {author} {\bibinfo {author} {\bibfnamefont {N.~C.}\ \bibnamefont
  {Keim}}, \bibinfo {author} {\bibfnamefont {J.~D.}\ \bibnamefont {Paulsen}},
  \bibinfo {author} {\bibfnamefont {Z.}~\bibnamefont {Zeravcic}}, \bibinfo
  {author} {\bibfnamefont {S.}~\bibnamefont {Sastry}}, \ and\ \bibinfo {author}
  {\bibfnamefont {S.~R.}\ \bibnamefont {Nagel}},\ }\href {\doibase
  10.1103/RevModPhys.91.035002} {\bibfield  {journal} {\bibinfo  {journal}
  {Rev. Mod. Phys.}\ }\textbf {\bibinfo {volume} {91}},\ \bibinfo {pages}
  {035002} (\bibinfo {year} {2019})}\BibitemShut {NoStop}%
\bibitem [{\citenamefont {Charbonneau}\ and\ \citenamefont
  {Morse}(2021)}]{morse21}%
  \BibitemOpen
  \bibfield  {author} {\bibinfo {author} {\bibfnamefont {P.}~\bibnamefont
  {Charbonneau}}\ and\ \bibinfo {author} {\bibfnamefont {P.~K.}\ \bibnamefont
  {Morse}},\ }\href {\doibase 10.1103/PhysRevLett.126.088001} {\bibfield
  {journal} {\bibinfo  {journal} {Phys. Rev. Lett.}\ }\textbf {\bibinfo
  {volume} {126}},\ \bibinfo {pages} {088001} (\bibinfo {year}
  {2021})}\BibitemShut {NoStop}%
\bibitem [{\citenamefont {Pashine}\ \emph {et~al.}(2019)\citenamefont
  {Pashine}, \citenamefont {Hexner}, \citenamefont {Liu},\ and\ \citenamefont
  {Nagel}}]{nagel19}%
  \BibitemOpen
  \bibfield  {author} {\bibinfo {author} {\bibfnamefont {N.}~\bibnamefont
  {Pashine}}, \bibinfo {author} {\bibfnamefont {D.}~\bibnamefont {Hexner}},
  \bibinfo {author} {\bibfnamefont {A.~J.}\ \bibnamefont {Liu}}, \ and\
  \bibinfo {author} {\bibfnamefont {S.~R.}\ \bibnamefont {Nagel}},\ }\href
  {\doibase 10.1126/sciadv.aax4215} {\bibfield  {journal} {\bibinfo  {journal}
  {Science Advances}\ }\textbf {\bibinfo {volume} {5}},\ \bibinfo {pages}
  {eaax4215} (\bibinfo {year} {2019})}\BibitemShut {NoStop}%
\bibitem [{\citenamefont {Hexner}\ \emph {et~al.}(2020)\citenamefont {Hexner},
  \citenamefont {Pashine}, \citenamefont {Liu},\ and\ \citenamefont
  {Nagel}}]{nagel20}%
  \BibitemOpen
  \bibfield  {author} {\bibinfo {author} {\bibfnamefont {D.}~\bibnamefont
  {Hexner}}, \bibinfo {author} {\bibfnamefont {N.}~\bibnamefont {Pashine}},
  \bibinfo {author} {\bibfnamefont {A.~J.}\ \bibnamefont {Liu}}, \ and\
  \bibinfo {author} {\bibfnamefont {S.~R.}\ \bibnamefont {Nagel}},\ }\href
  {\doibase 10.1103/PhysRevResearch.2.043231} {\bibfield  {journal} {\bibinfo
  {journal} {Phys. Rev. Research}\ }\textbf {\bibinfo {volume} {2}},\ \bibinfo
  {pages} {043231} (\bibinfo {year} {2020})}\BibitemShut {NoStop}%
\bibitem [{\citenamefont {Fiocco}\ \emph {et~al.}(2014)\citenamefont {Fiocco},
  \citenamefont {Foffi},\ and\ \citenamefont {Sastry}}]{sastry14}%
  \BibitemOpen
  \bibfield  {author} {\bibinfo {author} {\bibfnamefont {D.}~\bibnamefont
  {Fiocco}}, \bibinfo {author} {\bibfnamefont {G.}~\bibnamefont {Foffi}}, \
  and\ \bibinfo {author} {\bibfnamefont {S.}~\bibnamefont {Sastry}},\ }\href
  {\doibase 10.1103/PhysRevLett.112.025702} {\bibfield  {journal} {\bibinfo
  {journal} {Phys. Rev. Lett.}\ }\textbf {\bibinfo {volume} {112}},\ \bibinfo
  {pages} {025702} (\bibinfo {year} {2014})}\BibitemShut {NoStop}%
\bibitem [{\citenamefont {Lindeman}\ and\ \citenamefont
  {Nagel}(2021)}]{nagel21}%
  \BibitemOpen
  \bibfield  {author} {\bibinfo {author} {\bibfnamefont {C.~W.}\ \bibnamefont
  {Lindeman}}\ and\ \bibinfo {author} {\bibfnamefont {S.~R.}\ \bibnamefont
  {Nagel}},\ }\href {\doibase 10.1126/sciadv.abg7133} {\bibfield  {journal}
  {\bibinfo  {journal} {Science Advances}\ }\textbf {\bibinfo {volume} {7}},\
  \bibinfo {pages} {eabg7133} (\bibinfo {year} {2021})}\BibitemShut {NoStop}%
\bibitem [{\citenamefont {Scalliet}\ and\ \citenamefont
  {Berthier}(2019)}]{scalliet2019rejuvenation}%
  \BibitemOpen
  \bibfield  {author} {\bibinfo {author} {\bibfnamefont {C.}~\bibnamefont
  {Scalliet}}\ and\ \bibinfo {author} {\bibfnamefont {L.}~\bibnamefont
  {Berthier}},\ }\href@noop {} {\bibfield  {journal} {\bibinfo  {journal}
  {Physical review letters}\ }\textbf {\bibinfo {volume} {122}},\ \bibinfo
  {pages} {255502} (\bibinfo {year} {2019})}\BibitemShut {NoStop}%
\bibitem [{\citenamefont {Arceri}\ \emph {et~al.}(2020)\citenamefont {Arceri},
  \citenamefont {Landes}, \citenamefont {Berthier},\ and\ \citenamefont
  {Biroli}}]{arceri20}%
  \BibitemOpen
  \bibfield  {author} {\bibinfo {author} {\bibfnamefont {F.}~\bibnamefont
  {Arceri}}, \bibinfo {author} {\bibfnamefont {F.~P.}\ \bibnamefont {Landes}},
  \bibinfo {author} {\bibfnamefont {L.}~\bibnamefont {Berthier}}, \ and\
  \bibinfo {author} {\bibfnamefont {G.}~\bibnamefont {Biroli}},\ }\href@noop {}
  {\enquote {\bibinfo {title} {Glasses and aging: A statistical mechanics
  perspective},}\ } (\bibinfo {year} {2020}),\ \Eprint
  {http://arxiv.org/abs/2006.09725} {arXiv:2006.09725 [cond-mat.stat-mech]}
  \BibitemShut {NoStop}%
\bibitem [{\citenamefont {Kovacs}(1963)}]{kovacs1963glass}%
  \BibitemOpen
  \bibfield  {author} {\bibinfo {author} {\bibfnamefont {A.}~\bibnamefont
  {Kovacs}},\ }\href {https://link.springer.com/chapter/10.1007/BFb0050366}
  {\bibfield  {journal} {\bibinfo  {journal} {Adv. Polym. Sci}\ }\textbf
  {\bibinfo {volume} {3}},\ \bibinfo {pages} {394} (\bibinfo {year}
  {1963})}\BibitemShut {NoStop}%
\bibitem [{\citenamefont {Cugliandolo}\ \emph {et~al.}(2004)\citenamefont
  {Cugliandolo}, \citenamefont {Lozano},\ and\ \citenamefont
  {Lozza}}]{cugliandolo2004memory}%
  \BibitemOpen
  \bibfield  {author} {\bibinfo {author} {\bibfnamefont {L.}~\bibnamefont
  {Cugliandolo}}, \bibinfo {author} {\bibfnamefont {G.}~\bibnamefont {Lozano}},
  \ and\ \bibinfo {author} {\bibfnamefont {H.}~\bibnamefont {Lozza}},\ }\href
  {https://link.springer.com/article/10.1140/epjb/e2004-00298-2} {\bibfield
  {journal} {\bibinfo  {journal} {The European Physical Journal B-Condensed
  Matter and Complex Systems}\ }\textbf {\bibinfo {volume} {41}},\ \bibinfo
  {pages} {87} (\bibinfo {year} {2004})}\BibitemShut {NoStop}%
\bibitem [{\citenamefont {Bertin}\ \emph {et~al.}(2003)\citenamefont {Bertin},
  \citenamefont {Bouchaud}, \citenamefont {Drouffe},\ and\ \citenamefont
  {Godr{\`{e}}che}}]{bertin2003kovacs}%
  \BibitemOpen
  \bibfield  {author} {\bibinfo {author} {\bibfnamefont {E.~M.}\ \bibnamefont
  {Bertin}}, \bibinfo {author} {\bibfnamefont {J.-P.}\ \bibnamefont
  {Bouchaud}}, \bibinfo {author} {\bibfnamefont {J.-M.}\ \bibnamefont
  {Drouffe}}, \ and\ \bibinfo {author} {\bibfnamefont {C.}~\bibnamefont
  {Godr{\`{e}}che}},\ }\href {\doibase 10.1088/0305-4470/36/43/003} {\bibfield
  {journal} {\bibinfo  {journal} {Journal of Physics A: Mathematical and
  General}\ }\textbf {\bibinfo {volume} {36}},\ \bibinfo {pages} {10701}
  (\bibinfo {year} {2003})}\BibitemShut {NoStop}%
\bibitem [{\citenamefont {Song}\ \emph {et~al.}(2020)\citenamefont {Song},
  \citenamefont {Xu}, \citenamefont {Huo}, \citenamefont {Li}, \citenamefont
  {Wang}, \citenamefont {Ediger},\ and\ \citenamefont {Wang}}]{activation20}%
  \BibitemOpen
  \bibfield  {author} {\bibinfo {author} {\bibfnamefont {L.}~\bibnamefont
  {Song}}, \bibinfo {author} {\bibfnamefont {W.}~\bibnamefont {Xu}}, \bibinfo
  {author} {\bibfnamefont {J.}~\bibnamefont {Huo}}, \bibinfo {author}
  {\bibfnamefont {F.}~\bibnamefont {Li}}, \bibinfo {author} {\bibfnamefont
  {L.-M.}\ \bibnamefont {Wang}}, \bibinfo {author} {\bibfnamefont {M.~D.}\
  \bibnamefont {Ediger}}, \ and\ \bibinfo {author} {\bibfnamefont {J.-Q.}\
  \bibnamefont {Wang}},\ }\href {\doibase 10.1103/PhysRevLett.125.135501}
  {\bibfield  {journal} {\bibinfo  {journal} {Phys. Rev. Lett.}\ }\textbf
  {\bibinfo {volume} {125}},\ \bibinfo {pages} {135501} (\bibinfo {year}
  {2020})}\BibitemShut {NoStop}%
\bibitem [{\citenamefont {Prados}\ and\ \citenamefont
  {Brey}(2010)}]{prados2010kovacs}%
  \BibitemOpen
  \bibfield  {author} {\bibinfo {author} {\bibfnamefont {A.}~\bibnamefont
  {Prados}}\ and\ \bibinfo {author} {\bibfnamefont {J.~J.}\ \bibnamefont
  {Brey}},\ }\href {\doibase 10.1088/1742-5468/2010/02/p02009} {\bibfield
  {journal} {\bibinfo  {journal} {Journal of Statistical Mechanics: Theory and
  Experiment}\ }\textbf {\bibinfo {volume} {2010}},\ \bibinfo {pages} {P02009}
  (\bibinfo {year} {2010})}\BibitemShut {NoStop}%
\bibitem [{\citenamefont {Lahini}\ \emph {et~al.}(2017)\citenamefont {Lahini},
  \citenamefont {Gottesman}, \citenamefont {Amir},\ and\ \citenamefont
  {Rubinstein}}]{lahini2017nonmonotonic}%
  \BibitemOpen
  \bibfield  {author} {\bibinfo {author} {\bibfnamefont {Y.}~\bibnamefont
  {Lahini}}, \bibinfo {author} {\bibfnamefont {O.}~\bibnamefont {Gottesman}},
  \bibinfo {author} {\bibfnamefont {A.}~\bibnamefont {Amir}}, \ and\ \bibinfo
  {author} {\bibfnamefont {S.~M.}\ \bibnamefont {Rubinstein}},\ }\href
  {\doibase 10.1103/PhysRevLett.118.085501} {\bibfield  {journal} {\bibinfo
  {journal} {Phys. Rev. Lett.}\ }\textbf {\bibinfo {volume} {118}},\ \bibinfo
  {pages} {085501} (\bibinfo {year} {2017})}\BibitemShut {NoStop}%
\bibitem [{\citenamefont {K\"ursten}\ \emph {et~al.}(2017)\citenamefont
  {K\"ursten}, \citenamefont {Sushkov},\ and\ \citenamefont
  {Ihle}}]{kursten2017giant}%
  \BibitemOpen
  \bibfield  {author} {\bibinfo {author} {\bibfnamefont {R.}~\bibnamefont
  {K\"ursten}}, \bibinfo {author} {\bibfnamefont {V.}~\bibnamefont {Sushkov}},
  \ and\ \bibinfo {author} {\bibfnamefont {T.}~\bibnamefont {Ihle}},\ }\href
  {\doibase 10.1103/PhysRevLett.119.188001} {\bibfield  {journal} {\bibinfo
  {journal} {Phys. Rev. Lett.}\ }\textbf {\bibinfo {volume} {119}},\ \bibinfo
  {pages} {188001} (\bibinfo {year} {2017})}\BibitemShut {NoStop}%
\bibitem [{\citenamefont {Dillavou}\ and\ \citenamefont
  {Rubinstein}(2018)}]{rubinstein2018nonmonotonic}%
  \BibitemOpen
  \bibfield  {author} {\bibinfo {author} {\bibfnamefont {S.}~\bibnamefont
  {Dillavou}}\ and\ \bibinfo {author} {\bibfnamefont {S.~M.}\ \bibnamefont
  {Rubinstein}},\ }\href {\doibase 10.1103/PhysRevLett.120.224101} {\bibfield
  {journal} {\bibinfo  {journal} {Phys. Rev. Lett.}\ }\textbf {\bibinfo
  {volume} {120}},\ \bibinfo {pages} {224101} (\bibinfo {year}
  {2018})}\BibitemShut {NoStop}%
\bibitem [{\citenamefont {He}\ \emph {et~al.}(2019)\citenamefont {He},
  \citenamefont {Jiang}, \citenamefont {Chen}, \citenamefont {Liu},
  \citenamefont {Fan},\ and\ \citenamefont {Jiang}}]{he2019}%
  \BibitemOpen
  \bibfield  {author} {\bibinfo {author} {\bibfnamefont {Y.}~\bibnamefont
  {He}}, \bibinfo {author} {\bibfnamefont {D.}~\bibnamefont {Jiang}}, \bibinfo
  {author} {\bibfnamefont {J.}~\bibnamefont {Chen}}, \bibinfo {author}
  {\bibfnamefont {R.}~\bibnamefont {Liu}}, \bibinfo {author} {\bibfnamefont
  {J.}~\bibnamefont {Fan}}, \ and\ \bibinfo {author} {\bibfnamefont
  {X.}~\bibnamefont {Jiang}},\ }\href {\doibase 10.1007/s00603-018-1718-4}
  {\bibfield  {journal} {\bibinfo  {journal} {Rock Mechanics and Rock
  Engineering}\ }\textbf {\bibinfo {volume} {52}},\ \bibinfo {pages} {2471}
  (\bibinfo {year} {2019})}\BibitemShut {NoStop}%
\bibitem [{\citenamefont {Sánchez-Rey}\ and\ \citenamefont
  {Prados}(2020)}]{prados20}%
  \BibitemOpen
  \bibfield  {author} {\bibinfo {author} {\bibfnamefont {B.}~\bibnamefont
  {Sánchez-Rey}}\ and\ \bibinfo {author} {\bibfnamefont {A.}~\bibnamefont
  {Prados}},\ }\href@noop {} {\enquote {\bibinfo {title} {Linear response in
  the uniformly heated granular gas},}\ } (\bibinfo {year} {2020}),\ \Eprint
  {http://arxiv.org/abs/2010.10196} {arXiv:2010.10196 [cond-mat.stat-mech]}
  \BibitemShut {NoStop}%
\bibitem [{\citenamefont {Murphy}\ \emph {et~al.}(2020)\citenamefont {Murphy},
  \citenamefont {Kruppe},\ and\ \citenamefont {Jaeger}}]{murphy2020memory}%
  \BibitemOpen
  \bibfield  {author} {\bibinfo {author} {\bibfnamefont {K.~A.}\ \bibnamefont
  {Murphy}}, \bibinfo {author} {\bibfnamefont {J.~W.}\ \bibnamefont {Kruppe}},
  \ and\ \bibinfo {author} {\bibfnamefont {H.~M.}\ \bibnamefont {Jaeger}},\
  }\href {\doibase 10.1103/PhysRevLett.124.168002} {\bibfield  {journal}
  {\bibinfo  {journal} {Phys. Rev. Lett.}\ }\textbf {\bibinfo {volume} {124}},\
  \bibinfo {pages} {168002} (\bibinfo {year} {2020})}\BibitemShut {NoStop}%
\bibitem [{\citenamefont {Morgan}\ \emph {et~al.}(2020)\citenamefont {Morgan},
  \citenamefont {Avinery}, \citenamefont {Rahamim}, \citenamefont {Beck},\ and\
  \citenamefont {Saleh}}]{omar20}%
  \BibitemOpen
  \bibfield  {author} {\bibinfo {author} {\bibfnamefont {I.~L.}\ \bibnamefont
  {Morgan}}, \bibinfo {author} {\bibfnamefont {R.}~\bibnamefont {Avinery}},
  \bibinfo {author} {\bibfnamefont {G.}~\bibnamefont {Rahamim}}, \bibinfo
  {author} {\bibfnamefont {R.}~\bibnamefont {Beck}}, \ and\ \bibinfo {author}
  {\bibfnamefont {O.~A.}\ \bibnamefont {Saleh}},\ }\href {\doibase
  10.1103/PhysRevLett.125.058001} {\bibfield  {journal} {\bibinfo  {journal}
  {Phys. Rev. Lett.}\ }\textbf {\bibinfo {volume} {125}},\ \bibinfo {pages}
  {058001} (\bibinfo {year} {2020})}\BibitemShut {NoStop}%
\bibitem [{\citenamefont {Amir}\ \emph {et~al.}(2012)\citenamefont {Amir},
  \citenamefont {Oreg},\ and\ \citenamefont {Imry}}]{amir2012relaxations}%
  \BibitemOpen
  \bibfield  {author} {\bibinfo {author} {\bibfnamefont {A.}~\bibnamefont
  {Amir}}, \bibinfo {author} {\bibfnamefont {Y.}~\bibnamefont {Oreg}}, \ and\
  \bibinfo {author} {\bibfnamefont {Y.}~\bibnamefont {Imry}},\ }\href
  {https://doi.org/10.1073/pnas.1120147109} {\bibfield  {journal} {\bibinfo
  {journal} {Proceedings of the National Academy of Sciences}\ }\textbf
  {\bibinfo {volume} {109}},\ \bibinfo {pages} {1850} (\bibinfo {year}
  {2012})}\BibitemShut {NoStop}%
\bibitem [{\citenamefont {Chacko}\ \emph {et~al.}(2019)\citenamefont {Chacko},
  \citenamefont {Sollich},\ and\ \citenamefont {Fielding}}]{chacko19}%
  \BibitemOpen
  \bibfield  {author} {\bibinfo {author} {\bibfnamefont {R.~N.}\ \bibnamefont
  {Chacko}}, \bibinfo {author} {\bibfnamefont {P.}~\bibnamefont {Sollich}}, \
  and\ \bibinfo {author} {\bibfnamefont {S.~M.}\ \bibnamefont {Fielding}},\
  }\href {\doibase 10.1103/PhysRevLett.123.108001} {\bibfield  {journal}
  {\bibinfo  {journal} {Phys. Rev. Lett.}\ }\textbf {\bibinfo {volume} {123}},\
  \bibinfo {pages} {108001} (\bibinfo {year} {2019})}\BibitemShut {NoStop}%
\bibitem [{\citenamefont {Plata}\ and\ \citenamefont
  {Prados}(2017)}]{plata2017kovacs}%
  \BibitemOpen
  \bibfield  {author} {\bibinfo {author} {\bibfnamefont {C.~A.}\ \bibnamefont
  {Plata}}\ and\ \bibinfo {author} {\bibfnamefont {A.}~\bibnamefont {Prados}},\
  }\href {https://www.mdpi.com/1099-4300/19/10/539} {\bibfield  {journal}
  {\bibinfo  {journal} {Entropy}\ }\textbf {\bibinfo {volume} {19}},\ \bibinfo
  {pages} {539} (\bibinfo {year} {2017})}\BibitemShut {NoStop}%
\bibitem [{\citenamefont {Durian}(1995)}]{durianPRL1995}%
  \BibitemOpen
  \bibfield  {author} {\bibinfo {author} {\bibfnamefont {D.~J.}\ \bibnamefont
  {Durian}},\ }\href {\doibase 10.1103/PhysRevLett.75.4780} {\bibfield
  {journal} {\bibinfo  {journal} {Phys. Rev. Lett.}\ }\textbf {\bibinfo
  {volume} {75}},\ \bibinfo {pages} {4780} (\bibinfo {year}
  {1995})}\BibitemShut {NoStop}%
\bibitem [{\citenamefont {Durian}(1997)}]{DurianPRE1997}%
  \BibitemOpen
  \bibfield  {author} {\bibinfo {author} {\bibfnamefont {D.~J.}\ \bibnamefont
  {Durian}},\ }\href {\doibase 10.1103/PhysRevE.55.1739} {\bibfield  {journal}
  {\bibinfo  {journal} {Phys. Rev. E}\ }\textbf {\bibinfo {volume} {55}},\
  \bibinfo {pages} {1739} (\bibinfo {year} {1997})}\BibitemShut {NoStop}%
\bibitem [{\citenamefont {B{\"o}hmer}\ \emph {et~al.}(1995)\citenamefont
  {B{\"o}hmer}, \citenamefont {Schiener}, \citenamefont {Hemberger},\ and\
  \citenamefont {Chamberlin}}]{bohmer1995pulsed}%
  \BibitemOpen
  \bibfield  {author} {\bibinfo {author} {\bibfnamefont {R.}~\bibnamefont
  {B{\"o}hmer}}, \bibinfo {author} {\bibfnamefont {B.}~\bibnamefont
  {Schiener}}, \bibinfo {author} {\bibfnamefont {J.}~\bibnamefont {Hemberger}},
  \ and\ \bibinfo {author} {\bibfnamefont {R.}~\bibnamefont {Chamberlin}},\
  }\href {https://doi.org/10.1007/s002570050015} {\bibfield  {journal}
  {\bibinfo  {journal} {Zeitschrift f{\"u}r Physik B Condensed Matter}\
  }\textbf {\bibinfo {volume} {99}},\ \bibinfo {pages} {91} (\bibinfo {year}
  {1995})}\BibitemShut {NoStop}%
\bibitem [{\citenamefont {Marconi}\ \emph {et~al.}(2008)\citenamefont
  {Marconi}, \citenamefont {Puglisi}, \citenamefont {Rondoni},\ and\
  \citenamefont {Vulpiani}}]{marconi2008fluctuation}%
  \BibitemOpen
  \bibfield  {author} {\bibinfo {author} {\bibfnamefont {U.~M.~B.}\
  \bibnamefont {Marconi}}, \bibinfo {author} {\bibfnamefont {A.}~\bibnamefont
  {Puglisi}}, \bibinfo {author} {\bibfnamefont {L.}~\bibnamefont {Rondoni}}, \
  and\ \bibinfo {author} {\bibfnamefont {A.}~\bibnamefont {Vulpiani}},\ }\href
  {https://www.sciencedirect.com/science/article/abs/pii/S0370157308000768}
  {\bibfield  {journal} {\bibinfo  {journal} {Physics reports}\ }\textbf
  {\bibinfo {volume} {461}},\ \bibinfo {pages} {111} (\bibinfo {year}
  {2008})}\BibitemShut {NoStop}%
\bibitem [{\citenamefont {Diezemann}\ and\ \citenamefont
  {Heuer}(2011)}]{diezemann2011memory}%
  \BibitemOpen
  \bibfield  {author} {\bibinfo {author} {\bibfnamefont {G.}~\bibnamefont
  {Diezemann}}\ and\ \bibinfo {author} {\bibfnamefont {A.}~\bibnamefont
  {Heuer}},\ }\href {\doibase 10.1103/PhysRevE.83.031505} {\bibfield  {journal}
  {\bibinfo  {journal} {Phys. Rev. E}\ }\textbf {\bibinfo {volume} {83}},\
  \bibinfo {pages} {031505} (\bibinfo {year} {2011})}\BibitemShut {NoStop}%
\bibitem [{\citenamefont {Bedeaux}\ \emph {et~al.}(1971)\citenamefont
  {Bedeaux}, \citenamefont {Milosevic},\ and\ \citenamefont
  {Paul}}]{bedeaux1971linear}%
  \BibitemOpen
  \bibfield  {author} {\bibinfo {author} {\bibfnamefont {D.}~\bibnamefont
  {Bedeaux}}, \bibinfo {author} {\bibfnamefont {S.}~\bibnamefont {Milosevic}},
  \ and\ \bibinfo {author} {\bibfnamefont {G.}~\bibnamefont {Paul}},\ }\href
  {https://link.springer.com/article/10.1007/BF01012186} {\bibfield  {journal}
  {\bibinfo  {journal} {Journal of Statistical Physics}\ }\textbf {\bibinfo
  {volume} {3}},\ \bibinfo {pages} {39} (\bibinfo {year} {1971})}\BibitemShut
  {NoStop}%
\bibitem [{\citenamefont {Monthus}\ and\ \citenamefont
  {Bouchaud}(1996)}]{monthus1996models}%
  \BibitemOpen
  \bibfield  {author} {\bibinfo {author} {\bibfnamefont {C.}~\bibnamefont
  {Monthus}}\ and\ \bibinfo {author} {\bibfnamefont {J.-P.}\ \bibnamefont
  {Bouchaud}},\ }\href
  {https://iopscience.iop.org/article/10.1088/0305-4470/29/14/012} {\bibfield
  {journal} {\bibinfo  {journal} {Journal of Physics A: Mathematical and
  General}\ }\textbf {\bibinfo {volume} {29}},\ \bibinfo {pages} {3847}
  (\bibinfo {year} {1996})}\BibitemShut {NoStop}%
\bibitem [{\citenamefont {da~Mata}\ and\ \citenamefont
  {Pastor-Satorras}(2015)}]{da2015slow}%
  \BibitemOpen
  \bibfield  {author} {\bibinfo {author} {\bibfnamefont {A.~S.}\ \bibnamefont
  {da~Mata}}\ and\ \bibinfo {author} {\bibfnamefont {R.}~\bibnamefont
  {Pastor-Satorras}},\ }\href {https://doi.org/10.1140/epjb/e2014-50801-1}
  {\bibfield  {journal} {\bibinfo  {journal} {The European Physical Journal B}\
  }\textbf {\bibinfo {volume} {88}},\ \bibinfo {pages} {1} (\bibinfo {year}
  {2015})}\BibitemShut {NoStop}%
\bibitem [{\citenamefont {Bouchaud}\ \emph {et~al.}()\citenamefont {Bouchaud},
  \citenamefont {Cugliandolo}, \citenamefont {Kurchan},\ and\ \citenamefont
  {M\'ezard}}]{leticia97}%
  \BibitemOpen
  \bibfield  {author} {\bibinfo {author} {\bibfnamefont {J.~P.}\ \bibnamefont
  {Bouchaud}}, \bibinfo {author} {\bibfnamefont {L.~F.}\ \bibnamefont
  {Cugliandolo}}, \bibinfo {author} {\bibfnamefont {J.}~\bibnamefont
  {Kurchan}}, \ and\ \bibinfo {author} {\bibfnamefont {M.}~\bibnamefont
  {M\'ezard}},\ }\enquote {\bibinfo {title} {Out of equilibrium dynamics in
  spin-glasses and other glassy systems},}\ in\ \href {\doibase
  10.1142/9789812819437_0006} {\emph {\bibinfo {booktitle} {Spin Glasses and
  Random Fields}}},\ pp.\ \bibinfo {pages} {161--223}\BibitemShut {NoStop}%
\end{thebibliography}%
\bibliographystyle{ieeetr}

\end{document}